# Langevin equations for reaction-diffusion processes

Federico Benitez,[1,2] Charlie Duclut,[3] Hugues Chaté,[4,5,3] Bertrand Delamotte,[3] Ivan Dornic,[4,3] and Miguel A. Muñoz[6]

[1] Max Planck Institute for Solid State Research, Heisenbergstr. 1, 70569 Stuttgart, Germany
[2] Physikalisches Institut, Universität Bern, Sidlerstr. 5, 3012 Bern, Switzerland
[3] Laboratoire de Physique Théorique de la Matière Condensée,
CNRS UMR7600, UPMC-Sorbonne Universités, 75252 Paris Cedex 05, France
[4] Service de Physique de l'Etat Condensé, CEA, CNRS,
Université Paris-Saclay, CEA-Saclay, 91191 Gif-sur-Yvette, France
[5] Beijing Computational Science Research Center, Beijing 100094, China
[6] Instituto de Física Teórica y Computacional Carlos I,
Facultad de Ciencias, Universidad de Granada, 18071 Granada, Spain

(Dated: September 4, 2016)

For reaction-diffusion processes with at most bimolecular reactants, we derive well-behaved, numerically-tractable, exact Langevin equations that govern a stochastic variable related to the response field in field theory. Using *duality* relations, we show how the particle number and other quantities of interest can be computed. Our work clarifies long-standing conceptual issues encountered in field theoretical approaches and paves the way to systematic numerical and theoretical analyses of reaction-diffusion problems.

An important challenge in many areas of science is to reliably derive Langevin equations (LEs) governing the dynamics of relevant degrees of freedom at mesoscopic or coarse-grained scales. LEs are not only useful numerically, saving one from dealing with unessential details, but they also constitute the usual starting point for analytic approaches such as renormalization group analyses [1]. Not surprisingly, there is no systematic method to derive LEs from first principles, given that they are supposed to resum the complicated effects of different "forces" acting on the system at microscopic scales. As a matter of fact, LEs are often built from the underlying "mean-field" deterministic dynamics to which a noise term is added (see, e.g. [2, 3]). Several attempts to go beyond such a heuristic approach have been made in the past. Most use more or less sophisticated approximations to derive an equation for a "density" field [4–7]. The most popular ones are van Kampen's system size expansion [8] and Gillespie's projection method [9] (equivalent to a second-order truncation of the Kramers-Moyal equation), both of which rely on the presence of large numbers of particles. Although powerful within their range of validity, these approximations fail for spatially-extended systems displaying empty or sparsely populated regions.

Reaction-diffusion (RD) processes, which represent a wealth of phenomena in physics, chemistry, biology, population genetics, and even linguistics [10–13], play a particular role in this context in two respects: (i) they often exhibit empty domains and transition to absorbing states with strong fluctuations, and thus stand out of the validity range of popular methods to write LEs; (ii) their microscopic dynamics can be described exactly by a master equation, an object that can only be used directly for small systems but constitutes the starting point of field theoretical approaches with which, in turn, LEs can be *formally* associated. This procedure, using the Doi-Peliti formalism, [14–20] does not require large numbers of particles, and it would then appear that reliable LEs can be written for RD processes even in sparse regimes. However, as we show below, the LEs obtained often appear "paradoxical", the fields they govern may be difficult to interpret, and, worse, the path followed to derive them suffers from fundamental difficulties. Indeed, it often involves illegitimate steps, a consequence of which can be the advent of imaginary noise in equations supposed to describe real fields [21, 22]. A famous example of these inconsistencies is the simple, yet puzzling, pure annihilation case where particles from a single species diffuse and annihilate by pair upon encounter [19, 20, 22–25].

In this Letter, changing perspective, we present a systematic and exact derivation of LEs for RD processes that does *not* rely on field-theoretic tools, but rather stems directly from the master equation. These LEs do *not* govern a density field, but using *duality* considerations, a somewhat standard technique for stochastic processes and interacting particle systems [26–33], we show how the particle number and all its moments and correlations are computable from them. We also show how to relate our LEs to field theoretical approaches, resolving all previous paradoxes, and clarifying the meaning of the stochastic variable they govern. On top of the resolution of these long-standing conceptual issues, we show that this approach provides us with new practical and theoretical tools for computing observables of RD processes.

We consider the class of single-species RD processes involving all possible reactions of the form $A \xrightarrow{\alpha_p} pA$ ($p$ arbitrary), combined with any mixture of pairwise annihilation/coagulation $2A \xrightarrow{\beta_q} \emptyset$ and/or $A$ (with respective rates $\beta_0, \beta_1$). This set of reactions — comprising at most bimolecular reactants — already encompasses most of the physically interesting cases. For the sake of notational simplicity, we deal with the zero-dimensional case in what follows, since by considering diffusion as the nearest-neighbor hopping reaction $A\emptyset \overset{D}{\leftrightarrow} \emptyset A$ the generalization to $d$ dimensions is straightforward (see [34] for a detailed treatment). We define $N(t)$ as the stochastic variable representing the number of particles in the system at time $t$. The master equation that describes the evolution of the probabilities $P_n(t) \equiv \mathrm{Prob}(N(t) = n)$ reads

$$\partial_t P_n(t) = \sum_m L_{nm} P_m(t) \qquad (1)$$

where the elements of the rate transition matrix $L$ are $L_{nm} = \sum_p \alpha_p m (\delta_{n+1-p,m} - \delta_{m,n}) + \sum_q \beta_q m (m - 1)(\delta_{n+2-q,m} - \delta_{m,n})$. Introducing the probability-generating function $G(z,t) = \sum_{n=0}^{\infty} z^n P_n(t)$, Eq. (1) can be subsumed as an evolution equation for $G(z,t)$:

$$\partial_t G(z,t) = \mathcal{L}_z \, G(z,t) \qquad (2)$$

which involves a second order evolution operator $\mathcal{L}_z = \mathcal{L}(z, \partial_z) = A(z)\partial_z + B(z)\partial_z^2$, determined by the reaction rates through:

$$A(z) = \sum_p \alpha_p (z^p - z), \ \ B(z) = \sum_q \beta_q (z^q - z^2). \qquad (3)$$

Note that for the reactions we deal with, the coefficient of the second derivative reduces to a second order polynomial in $z$:

$$B(z) = (\beta_0 + \beta_1)(z - \ell)(1 - z), \ \ \ell \equiv -\beta_0/(\beta_0 + \beta_1). \qquad (4)$$

The consideration of $G$ — a well-defined analytic function for $z$ in the complex unit disk $|z| \leq 1$ — is a somewhat standard tool [35]. Besides the probability normalization encoded through $G(z = 1, t) = 1$, derivatives of $G(z, t)$ with respect to $z$ evaluated at $z = 1$ give access (in principle) to the (factorial) moments of the number of particles. Some recent works in the literature have also shown how one can extract rare-events statistics from the behavior of $G(z,t)$ near $z = 0$ [36].

The second order operator $\mathcal{L}_z$ governing the evolution of $G$ is *not* a Fokker-Planck operator that could be associated with the number of particles $N(t)$. However, $\mathcal{L}_z^\dagger$, the Hermitian conjugate of $\mathcal{L}_z$, is the Fokker-Planck operator for a stochastic variable $Z(t)$ obeying the Itō (prepoint) LE:

$$\dot{Z}(t) = A(Z(t)) + \sqrt{2B(Z(t))} \, \eta(t), \qquad (5)$$

where $\eta(t)$ is a zero mean unit variance Gaussian white noise. Crucially, due to the form of the drift and diffusion functions $A(z), B(z)$ given in Eqs. (3) and (4), the stochastic variable $Z(t)$ always remains (if initially so) in the bounded *real* interval $[\ell, 1]$ where $\sqrt{B(z)} \geq 0$. We show in the following that the determination of the statistics of $Z(t)$ through the knowledge of $G$ suffices to extract 'all the physics' of interest for the original RD process.

With its multiplicative noise and its two square-root barriers at $\ell$ and 1, Eq. (5) resembles phenomenological LEs usually considered for nonequilibrium phase transitions with symmetric absorbing states [37–43]. However, the LE (5) is *exact* and is $Z(t)$ *not* a density: we show below that $Z(t)$ is in fact closely related to the (time-reversed) response field used in field theory and explain how useful quantities can be computed from Eq. (5).

The evolution of $Z(t)$ stops when it has reached the absorbing barrier located at 1 (whose fixed location can be traced back to the probability conservation). This implies the existence of a delta-peak term at $z = 1$ in the probability distribution $p(z, t)$ of $Z(t)$ and, depending on the values of the $\alpha_p$'s, whenever (the otherwise always nonnegative) $A(\ell)$ vanishes, a second delta-peak appears at $z = \ell$. Thus, the general form of $p(z, t)$ reads:

$$p(z,t) = p_c(z,t) + q_1(t)\delta(1 - z) + q_\ell(t)\delta(\ell - z) \qquad (6)$$

where $p_c(z, t)$ is the continuous part of the distribution and $q_1, q_\ell$ the weights at the boundaries. Note finally that efficient and accurate methods dealing properly with multiplicative square-root noise exist [37, 38, 44–46], so that LE (5) can be used numerically (even in the spatially-extended case), and thereby the obtained statistics of $Z(t)$ faithfully reconstructs Eq. (6).

We now derive the fundamental relation that allows us to compute quantities for the original RD process. Using Itō calculus in Eq. (5) [30, 35] or, alternatively, its associated Fokker-Planck equation, one can show that

$$\partial_t \langle Z(t)^n \rangle_{\mathrm{LE}} = \sum_m (L^T)_{nm} \langle Z(t)^m \rangle_{\mathrm{LE}} \qquad (7)$$

where $L^T$ is the transpose matrix of $L$ (see Eq. (1)), and $\langle \cdot \rangle_{\mathrm{LE}}$ denotes averaging over the noise $\eta(t)$ of Eq. (5), that is with the distribution $p(z, t)$. Combining Eqs. (1) and (7), it is now easy to show that, for any fixed time $t$, the quantity $\sum_n P_n(t-s) \langle Z(s)^n \rangle_{\mathrm{LE}}$ is independent of $s$ ($0 \leq s \leq t$). Evaluating it at $s = t$ and $s = 0$, we find the exact *duality* relation [47]

$$\int_\ell^1 \mathrm{d}z \sum_{n=0}^{\infty} p(z,t) P_n(0) z^n = \int_\ell^1 \mathrm{d}z \sum_{n=0}^{\infty} p(z,0) P_n(t) z^n \qquad (8)$$

that also reads: $\langle \langle Z(t)^{N(0)} \rangle_{\mathrm{LE}} \rangle_{\mathrm{RD}} = \langle \langle Z(0)^{N(t)} \rangle_{\mathrm{LE}} \rangle_{\mathrm{RD}}$ where $\langle \cdot \rangle_{\mathrm{RD}}$ has to be understood as an averaging over the RD process. (This relation generalizes an analogous one derived by Doering *et al.* for the reversible coagulation-decoagulation process $A \rightleftharpoons 2A$ [30].)

Using Eq. (8) and choosing properly initial conditions of the LE, one easily computes quantities of the RD process such as the survival probability and the moments of the probability distribution $P_n(t)$. The survival probability $P_{\mathrm{surv}}^{(m)}$ is defined as the probability that, starting at $t = 0$ with $m > 0$ particles, at least one particle survives at time $t$: $P_{\mathrm{surv}}^{(m)}(t) = 1 - P_0(t)$. Using Eq. (8) with

$p(z, 0) = \delta(z)$ and $P_n(0) = \delta_{mn}$, we obtain

$$1 - P_{\mathrm{surv}}^{(m)}(t) = \int_{\ell}^{1} \mathrm{d}z \, p(z, t | 0, 0) z^m = \langle Z(t)^m \rangle_{\mathrm{LE}} \quad (9)$$

where $p(z, t | z_0, 0)$ is the conditional transition probability of the LE with $p(z, t = 0) = \delta(z - z_0)$ as initial condition and the $m$th-order moment is a readily measurable quantity in a LE simulation. Similarly, the moments of the RD process can be derived from $G(z, t)$ using $p(z, 0) = \delta(z - z_0)$ as initial condition; Eq. (8) yields [48]:

$$G(z_0, t) = \int_{\ell}^{1} \mathrm{d}z \sum_{n=0}^{\infty} p(z, t | z_0, 0) P_n(0) z^n. \quad (10)$$

Differentiating $G(z_0, t)$ with respect to $z_0$ and evaluating it at $z_0 = 1$ yields the (factorial) moments of the RD process. For instance, the average particle number reads

$$\langle N(t) \rangle = \partial_{z_0} \int_{\ell}^{1} \mathrm{d}z \sum_{n} p(z, t | z_0, 0) P_n(0) z^n \Big|_{z_0 = 1}. \quad (11)$$

All formulas above can be easily generalized to the spatially extended case in the presence of diffusion. For instance, putting $m$ particles at one site $i$ and 0 elsewhere in the RD process and choosing $Z_j(0) = 0$ for all sites $j$ for the LE, Eq. (9) is replaced by: $P_{\mathrm{surv}}^{(m, \, i)}(t) = 1 - \langle Z_i(t)^m \rangle_{\mathrm{LE}}$ [49].

We now make contact with the field-theoretic approaches alluded to above. In addition to clarifying the situation there, this elucidates the physical meaning of the stochastic variable $Z$ governing Eq. (5), and also reveals how correlation functions at different times and response functions can be calculated within our framework.

We first recall the main features of the Doi-Peliti formalism from which follow the field theories associated with RD processes (see, e.g., [12, 15, 17, 18]). A (state) vector $|P(t)\rangle$ is associated with the set of probabilities $\{P_n(t)\}$. This vector belongs to a Hilbert space spanned by the "occupation number" vectors $\{|n\rangle\}$ and reads: $|P(t)\rangle = \sum_{n=0}^{\infty} P_n(t) |n\rangle$. The vector $|n\rangle$ is an eigenvector with eigenvalue $n$ of the "number" operator $\hat{N} = a^{\dagger} a$ where $a$ and $a^{\dagger}$ are annihilation and creation operators satisfying $[a, a^{\dagger}] = 1$, $a|0\rangle = 0$, $a|n\rangle = n|n-1\rangle$, and $a^{\dagger}|n\rangle = |n+1\rangle$. The scalar product is chosen such that $\langle m|n\rangle = n! \delta_{mn}$ and $a^{\dagger}$ is the Hermitian conjugate of $a$. With any complex number $\phi$ is associated a *coherent state* $|\phi\rangle$ defined by the relation $|\phi\rangle = \exp(\phi a^{\dagger})|0\rangle$. The probability generating function can then be written $G(z, t) = \langle z|P(t)\rangle$. The evolution of the $\{P_n(t)\}$ induces the evolution of the state vector: $\partial_t |P(t)\rangle = \mathcal{L}(a^{\dagger}, a) |P(t)\rangle$ where $\mathcal{L}$, written in its normal-ordered form, is the very same function as in Eq. (2). With the LE probability distribution $p(z, t)$ defined in Eq. (6), we also associate a vector:

$$|p(t)\rangle = \int_{\ell}^{1} \mathrm{d}z \, p_c(z, t) |z\rangle + q_1(t)|1\rangle + q_{\ell}(t)|\ell\rangle \quad (12)$$

where $|z\rangle$ is the coherent state with real eigenvalue $z \in [\ell, 1]$. From the evolution of $p(z, t)$ one checks that $\partial_t |p(t)\rangle = \mathcal{L}^{\dagger}(a^{\dagger}, a)|p(t)\rangle$ where $\mathcal{L}^{\dagger}(a^{\dagger}, a)$ is also normal-ordered. Using the resolution of the identity $2\pi \mathbb{I} = \iint_0^{\infty} \mathrm{d}z \, \mathrm{d}z' \exp(-izz') |iz\rangle \langle z'|$, one can show that Eq. (8) reads $\langle p(t)|P(0)\rangle = \langle p(0)|P(t)\rangle$ [32, 33]. Thus, within the Doi-Peliti formalism, the duality relation (Eq. (8)) is a direct consequence of the fact that $|P(t)\rangle$ evolves with $\mathcal{L}$ and $|p(t)\rangle$ with $\mathcal{L}^{\dagger}$.

The field theoretical approach to RD processes is based on a functional integral representation of the evolution operator $\exp(\mathcal{L}t)$ of the state vector $|P(t)\rangle$. Using the Trotter formula and introducing resolutions of the identity in terms of *complex-conjugate* coherent states one can derive the generating functional of correlation and response functions in the presence of real sources [50]:

$$\mathcal{Z}[J, \tilde{J}] = \int \mathcal{D}\phi \mathcal{D}\phi^* \, \mathrm{e}^{-S[\phi, \phi^*] + \int \mathrm{d}t (J\phi + \tilde{J}\phi^*)}, \quad (13)$$

with the action $S[\phi, \phi^*] = \int \mathrm{d}t \, [\phi^* \partial_t \phi - \mathcal{L}(\phi^*, \phi)]$, where $\phi$ and $\phi^*$ are complex-conjugate fields.

The usual derivation of a LE from this field theory goes as follows (see, e.g., [20, 25, 51]). After performing the shift $\phi^* \to \phi^* + 1$, the fields $\phi$ and $\phi^*$ are formally replaced in the action $S$ by a real field $\psi$, dubbed the "density" field, and an imaginary field $\tilde{\psi}$, called the response field. For binary reactions, $\mathcal{L}(\tilde{\psi} + 1, \psi)$ is at most quadratic in $\tilde{\psi}$, the term $\exp(\int \mathrm{d}t \, \tilde{\psi}^2 U(\psi))$ with $U(\psi) = \alpha_2 \psi - (\beta_0 + \beta_1)\psi^2$ in $\mathcal{Z}$ is formally written as a Gaussian integral $\int \mathcal{D}\eta \exp(-\int \mathrm{d}t \, [\eta^2/2 + \sqrt{2U(\psi)} \tilde{\psi}\eta])$ and the resulting argument of the exponential is thus linear in $\tilde{\psi}$. The functional integral on the imaginary field $\tilde{\psi}$ then yields:

$$\mathcal{Z}[J, \tilde{J}] = \int \mathcal{D}\psi \mathcal{D}\eta \, \mathcal{P}(\eta(t)) \delta(f(\psi, \eta, \tilde{J})) \, \mathrm{e}^{\int \mathrm{d}t \, J\psi} \quad (14)$$

where $\mathcal{P}(\eta(t)) = \exp(-\int \mathrm{d}t \, \eta(t)^2/2)$, $f = -\partial_t \psi + \alpha_2 \psi - (2\beta_0 + \beta_1)\psi^2 + \tilde{J} + \sqrt{2(\alpha_2 \psi - (\beta_0 + \beta_1)\psi^2)} \, \eta$, and $\delta(f(\psi, \eta, \tilde{J}))$ is a functional Dirac function. Written under this form, $\mathcal{Z}[J, \tilde{J}]$ is the generating functional of correlation functions derived from the LE: $f(\psi, \eta, \tilde{J}) = 0$ where $\eta(t)$ is interpreted as a Gaussian white noise and the derivation above follows the standard Martin-Siggia-Rose-De Dominicis-Janssen (MSRDJ) method in the reverse order [52–54]. For instance, for pure annihilation $(2A \to \emptyset)$, this yields the 'imaginary noise' LE:

$$\partial_t \psi = -2\beta_0 \psi^2 + i \sqrt{2\beta_0} \, \psi \, \eta. \quad (15)$$

The problem with this derivation is that it is purely formal. Although exact to all orders of perturbation theory [34], it is actually incorrect to trade the two complex-conjugate fields $\phi$ and $\phi^*$ for a real and an imaginary field in $\mathcal{Z}[J, \tilde{J}]$ since the resulting functional integral is in general no longer convergent at large fields (a fact that is immaterial within perturbation theory). Indeed, the leading term at large fields $-\beta_1\psi^2\tilde\psi^2$ has the wrong sign since $\tilde\psi$ is purely imaginary, contrarily to the original term $-\beta_1\phi^2\phi^{*2}$. The imaginary noise in Eq. (15) is a consequence of this formal and incorrect step. Notice that the derivation is performed assuming that $\psi(t)$ is real, while a field evolving according to Eq. (15) necessarily becomes complex.

To overcome the convergence problems discussed above and the fact that the action $S$ is not quadratic in $\phi^*$ when reactions $A \to pA$ with $p > 2$ are involved, we now provide a proper and general field-theoretical derivation of a LE by using contour deformations in the complex plane. In contrast with the usual procedure, $\phi$ is deformed into a purely imaginary variable $\psi$, and the conjugated field $\phi^*$ into a real variable $\tilde\psi \in [\ell, 1]$. A complete derivation for a particular RD process is presented in the Supp. Mat. [34]. In the spatially extended case where particles diffuse with rate $D$, the result reads:

$$\mathcal{Z}[J,\tilde J] = \int_\ell^1 \mathcal{D}\tilde\psi \int_{-i\infty}^{i\infty} \mathcal{D}\psi \, \mathrm{e}^{-S[\psi,\tilde\psi]+\int_{t,x}(J\psi+\tilde J\tilde\psi)}, \quad (16)$$

where the functional form of $S$ has not changed (it still involves the same function $\mathcal{L}$ as above, but its arguments are different). Finally, using the same step as discussed above (the MSRDJ formalism in the reverse order) consisting of: (i) rewriting the quadratic term in $\psi$ as a functional integral over a Gaussian field $\eta$ and (ii) integrating over the imaginary variable $\psi$, we obtain $\mathcal{Z}[J,\tilde J]$ under a form similar to (Eq. (14)) with $\psi$ replaced by $\tilde\psi$ and $f(\psi,\eta,\tilde J)$ replaced by $g(\tilde\psi,\eta,J) = \partial_t\tilde\psi + D\nabla^2\tilde\psi + A(\tilde\psi) + J + \sqrt{2B(\tilde\psi)}\,\eta$. Notice that the integration over $\psi$ leading to $\delta(g)$ requires an integration by parts in $S$ of the $\tilde\psi\partial_t\psi$ term that changes its sign. A change of $t$ into $-t$ is thus necessary to get the usual diffusion term as can be seen on the explicit expression of $g$ above. Therefore, the corresponding LE for $\tilde\psi$ runs backwards in time. Setting $J = 0$ and defining $Z(t) \equiv \tilde\psi(-t)$, we finally obtain exactly the same LE as Eq. (5).

The derivation above shows that the field $Z(t)$ of the LE (5) is closer in spirit to the (time-reversed) response-field than to the direct (density) field, and also that the source term $\int J\psi$ in $\mathcal{Z}[J,\tilde J]$ —which is convenient to derive correlation functions but may seem unphysical— has a simple meaning here: it appears as an external force in (5), which can then be used to calculate not only response functions but also all correlation functions. [55].

As a specific example of the method, we consider the reactions: $2A \xrightarrow{\beta_0} \emptyset$ and $A \xrightarrow{\alpha_2} 2A$, together with diffusion at rate $D$. This RD process is archetypical of the prominent directed percolation (DP) universality class [2, 3]. Through the Doi-Peliti formalism, the action is readily derived. Using the contour deformations described in

[34], and defining $Z(x,t) = \tilde\psi(x,-t)$ we obtain:

$$\partial_t Z = D\nabla^2 Z - \alpha_2 Z(1-Z) + \sqrt{2\beta_0(1-Z^2)}\,\eta \quad (17)$$

with $Z = Z(x,t) \in [-1,1]$ (as determined by the contour deformations) which is identical to Eq. (5) for this set of reactions (supplemented with diffusion). For DP, the action is quadratic both in $\phi$ and $\phi^*$ and another contour deformation [56] leads to a LE on the other field $\psi(x,t)$:

$$\partial_t\psi = D\nabla^2\psi + \psi(\alpha_2 - 2\beta_0\psi) + \sqrt{2\psi(\alpha_2 - \beta_0\psi)}\,\eta \quad (18)$$

with now $\psi = \psi(x,t) \in [0, \alpha_2/\beta_0]$. Note that Eqs. (17) and (18) are identical up to a simple linear change of variable, something referred to as "rapidity symmetry". Note however that Eq. (18) ceases to exist in the pure annihilation limit $\alpha_2 \to 0$ ("imaginary noise") [57]. In contrast, Eq. (17) remains well-behaved, and should be considered as the *correct* LE in such a case.

Another well-established universality class of nonequilibrium absorbing phase transitions corresponds to RD processes where the parity of the (local) number of particles is preserved by the reactions [18, 58]. Viewing the particles as domain walls between two symmetric absorbing states, the following phenomenological LE with two symmetric absorbing barriers was postulated in [39]:

$$\partial_t\varphi = D\nabla^2\varphi + (a\varphi - b\varphi^3)(1-\varphi^2) + c\sqrt{1-\varphi^2}\,\eta. \quad (19)$$

With hindsight, this LE turns out to be the exact (time-reversed) response-field LE for the RD process $A \xrightarrow{\alpha_3} 3A$, $A \xrightarrow{\alpha_5} 5A$, $2A \xrightarrow{\beta_0} \emptyset$, that indeed belongs to this class. Last but not least, it provides us with the microscopic parameters identification $a = -(\alpha_3 + \alpha_5)$, $b = \alpha_5$, and $c = \sqrt{2\beta_0}$.

To summarize, we derived exact LEs for RD processes with at most bimolecular reactants and showed how they can be used to calculate usual quantities of interest. Our work clarifies several misunderstandings that have haunted the related field-theoretical literature for decades: Why some RD processes lead to imaginary noise, why some can be written in terms of the so-called density field whereas others can only be written in terms of the response-field, and how keeping track of the evolution of the latter allows for computing correlation functions of the original RD process.

Beyond its obvious importance in physics, chemistry, and the many fields where problems can be explicitly formulated in the form of RD processes, our work may have particular impact for the numerous situations where equations similar to Eq. (5) have been written [29, 40–43, 59–64]. We hope our results will be put into practice and help strengthen and clarify the use of Langevin equations in various fields.

We thank Claude Aslangul for keen advice and help on analytic functions theory, Hélène Berthoumieux and Gilles Tarjus for useful discussions.

# Langevin equations for reaction-diffusion processes

Federico Benitez,[1,2] Charlie Duclut,[3] Hugues Chaté,[4,5,3] Bertrand Delamotte,[3] Ivan Dornic,[4] and Miguel A. Muñoz[6]

[1]*Max Planck Institute for Solid State Research, Heisenbergstr. 1, 70569 Stuttgart, Germany*
[2]*Physikalisches Institut, Universität Bern, Sidlerstr. 5, 3012 Bern, Switzerland*
[3]*Laboratoire de Physique Théorique de la Matière Condensée,*
*CNRS UMR7600, UPMC-Sorbonne Universités, 75252 Paris Cedex 05, France*
[4]*Service de Physique de l'Etat Condensé, CEA, CNRS,*
*Université Paris-Saclay, CEA-Saclay, 91191 Gif-sur-Yvette, France*
[5]*Beijing Computational Science Research Center, Beijing 100094, China*
[6]*Instituto de Física Teórica y Computacional Carlos I,*
*Facultad de Ciencias, Universidad de Granada, 18071 Granada, Spain*
(Dated: September 4, 2016)

**Fokker-Planck equation for the dual variable $Z(t)$ and evolution of its moments.**

The probability distribution $p(z,t)$ of the variable $Z(t)$ evolves according to the Fokker-Planck equation associated with the LE (4) of the main text. The Fokker-Planck operator is not $\mathcal{L}(z, \partial_z)$ defined in Eq.(3) but, as we show below, its hermitic conjugate (in function space) $\mathcal{L}^\dagger(z, \partial_z)$:

$$\partial_t p(z,t) = \mathcal{L}^\dagger(z, \partial_z) p(z,t). \tag{1}$$

The probability distribution $p(z,t)$ exhibits not only a continuous part but also delta-peaks at $z = \ell$ and $z = 1$, to enforce probability conservation. We start by presenting an elementary derivation of the evolution of the continuous part of the probability distribution, $p_c(z,t)$. By definition, $p_c(z,t) = \langle \delta(Z(t) - z) \rangle_{\text{LE}}$ with $\ell < z < 1$. A convenient way is to compute $\delta p_c = p_c(z, t + \delta t) - p_c(z,t)$ for small $\delta t$. Between $t$ and $t + \delta t$ and with Ito discretization, the variable $Z$ evolves according to

$$Z(t + \delta t) = Z(t) + A(Z(t))\delta t + \sqrt{2B(Z(t))}\,\delta W(t) \tag{2}$$

where $\delta W(t) = \eta(t)\delta t$ is the increment of the white noise which is faithfully represented (for $\delta t$ small enough) by a standard Gaussian normal variable with variance $\delta t$: $\delta W(t) = N(0, \delta t) = \sqrt{\delta t}N(0,1)$. Using an integral representation for the delta function, we obtain at order $\delta t$:

$$\begin{aligned}\delta p_c &= \int \frac{dk}{2\pi} \langle e^{ik(Z(t+\delta t)-z)} - e^{ik(Z(t)-z)} \rangle_{\text{LE}} \\ &= \delta t \int \frac{dk}{2\pi} \langle e^{ik(Z(t)-z)} \left[ikA(Z(t)) + k^2 B(Z(t))\right] \rangle_{\text{LE}}\end{aligned} \tag{3}$$

where the average $\langle \cdots \rangle_{\text{LE}}$ in the first line means an average over the noise histories up to time $t + \delta t$ while it is up to time $t$ in the second line. Notice that to go from the first to the second line we have used the Markov property $\langle \cdots \rangle_{\text{LE}(t+\delta t)} = \langle\langle \cdots \rangle_{\text{LE}(t)} \rangle_{\delta W(t)}$ to perform the average over the Gaussian random variable $N(0,1)$. Using $\langle A(Z(t))\delta(Z(t) - z)\rangle = A(z)p_c(z,t)$, we finally obtain:

$$\partial_t p_c(z,t) = -\partial_z(A(z)p_c(z,t)) + \partial_z^2(B(z)p_c(z,t)). \tag{4}$$

Given the form of the drift and diffusion coefficients $A(z), B(z)$, there is a non-zero probability current on the boundaries. To maintain probability conservation, it is therefore necessary to include these currents in the time-evolution of the total probability distribution $p(z,t)$, which reads

$$\partial_t p = \bar{\mathcal{L}}_z\, p_c(z,t) + j_1(t)\delta(1-z) - j_\ell(t)\delta(\ell-z) \tag{5}$$

with

$$\begin{aligned}\bar{\mathcal{L}}_z\, p_c(z,t) &= -\partial_z[A(z)p_c(z,t)] + \partial_z^2[B(z)p_c(z,t)] \\ j_z(t) &= A(z)p_c(z,t) - \partial_z[B(z)p_c(z,t)]\end{aligned} \tag{6}$$

This shows that the evolution of the weights $q_1(t)$ and $q_\ell(t)$ of the delta-peaks comes from the probability current $j_z(t)$ of the continuous part, evaluated at the boundaries. Equations (5),(6) yield the explicit form of $\mathcal{L}^\dagger(z, \partial_z)$ in Eq. (1).

Let us now derive the evolution equation for an arbitrary observable $\langle \mathcal{O}(Z(t)) \rangle_{\text{LE}}$. By definition, the stochastic variable $Z(t)$ takes the value $z$ at time $t$ with probability $p(z,t)$, and therefore

$$\frac{d}{dt}\langle \mathcal{O}(Z(t)) \rangle_{\text{LE}} = \int_\ell^1 dz\, \partial_t p(z,t)\, \mathcal{O}(z) \tag{7}$$

where the integral has to be understood as $\int_\ell^1 dz \cdots \equiv \int_{\ell^-}^{1^+} dz \cdots$ so that the contributions of the delta-peaks in Eq. (5) are taken into account. Substituting Eq. (6) in Eq. (7), and performing an integration by parts, one finds that the delta-peaks contribution are cancelled by the boundary terms, so that

$$\begin{aligned}\frac{d}{dt}\langle \mathcal{O}(Z(t)) \rangle_{\text{LE}} &= \int_\ell^1 dz\, p_c(z,t)\left[A(z)\partial_z + B(z)\partial_z^2\right]\mathcal{O}(z) \\ &= \int_\ell^1 dz\, p(z,t)\,\mathcal{L}_z\mathcal{O}(z) \tag{8}\\ &\equiv \langle \mathcal{L}_{Z(t)}\mathcal{O} \rangle_{\text{LE}} \tag{9}\end{aligned}$$

where the second equality holds because both $A(z)$ and $B(z)$ coefficients vanish when $z = 1$ and $z = \ell$.

Notice finally that the above general computations amount to check that one can define a scalar product $\prec \cdot \succ$ on the interval $[\ell, 1]$ (more precisely, a pairing between probability distributions and functions defined there) such that

$$\frac{\mathrm{d}}{\mathrm{d}t}\langle \mathcal{O}(Z(t))\rangle_{\mathrm{LE}} = \prec \mathcal{L}^\dagger[p] \cdot \mathcal{O} \succ = \prec p \cdot \mathcal{L}[\mathcal{O}] \succ \quad (10)$$

where in the last equality is it immaterial to use either $p$ or $p_c$. Equation (10) justifies why $\mathcal{L}^\dagger$ is called the hermitic conjugate (or adjoint) of $\mathcal{L}$.

To establish Eq. (6) of the main text, we now choose $\mathcal{O}(Z) = Z^n$, and use Eq. (7) as well as the specific form of $\mathcal{L}_z$ as given by Eq. (3). This immediately yields

$$\frac{\mathrm{d}}{\mathrm{d}t}\langle (Z(t))^n\rangle_{\mathrm{LE}} = \sum_p \alpha_p n(\langle Z(t)^{n+p-1}\rangle_{\mathrm{LE}} - \langle Z(t)^n\rangle_{\mathrm{LE}})$$
$$+ \sum_q \beta_q n(n-1)(\langle Z(t)^{n+q-2}\rangle_{\mathrm{LE}} - \langle Z(t)^n\rangle_{\mathrm{LE}}). \quad (11)$$

Recalling the expression of the rate transition matrix:

$$L_{nm} = \sum_p \alpha_p m(\delta_{n+1-p,m} - \delta_{m,n})$$
$$+ \sum_q \beta_q m(m-1)\left(\delta_{n+2-q,m} - \delta_{m,n}\right), \quad (12)$$

the moment duality formula (6) of the main text follows directly.

**Perturbative mapping between Doi-Peliti and MSRDJ formalisms**

We review the usual (zero-dimensional) perturbative argument showing that the functional integral written in terms of complex-conjugate (coherent states) fields $\phi$ and $\phi^*$ can be equivalently written in terms of a purely real field $\psi$ and a purely imaginary field $\tilde{\psi}$. The starting point is the following relations:

$$n!\,\delta_{mn} = \int_{-\infty}^{+\infty} \mathrm{d}x\, x^n \left(-\frac{d}{dx}\right)^m \delta(x) \quad (13)$$

$$= \frac{1}{2\pi}\int_{-\infty}^{+\infty} \mathrm{d}x\, \mathrm{d}y\, x^n (iy)^m \mathrm{e}^{-ixy}. \quad (14)$$

Now, working with a complex number $z$ we have:

$$n!\,\delta_{mn} = \frac{1}{\pi}\int \mathrm{d}z\, \mathrm{d}z^*\, z^n (z^*)^m \mathrm{e}^{-|z|^2} \quad (15)$$

which is easily proven by considering polar variables and where, by definition, $\mathrm{d}z\,\mathrm{d}z^* = \mathrm{d}(\mathrm{Re}z)\mathrm{d}(\mathrm{Im}z)$. Thus, up to a factor 2, we find that for this integral we obtain the same result if we consider as in Eq.(15) that $z$ and $z^*$ are complex conjugate variables or if we take as in Eq.(14) that $z$ is real ($z \to x$) and $z^*$ imaginary ($z \to iy$) and independent of $z$. It follows that for any Gaussian measure (denoted by an index 0):

$$\langle z^n (z^*)^m\rangle_0 = \langle x^n (iy)^m\rangle_0, \quad (16)$$

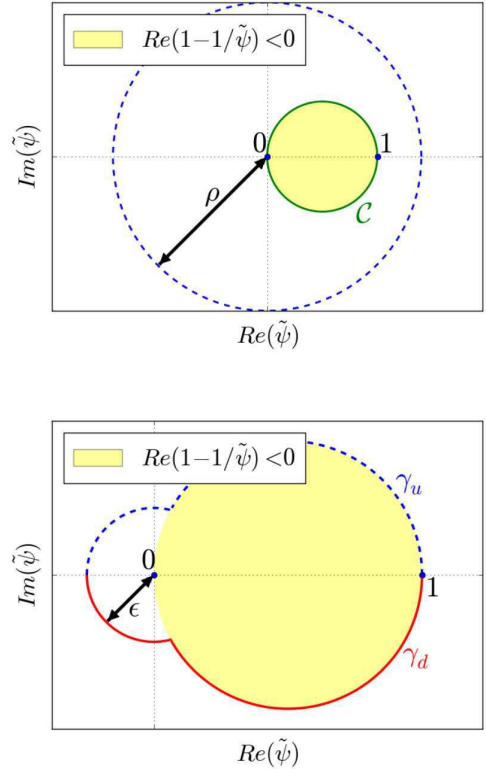

FIG. 1. (top) The initial integration circle $|\tilde{\psi}| = \rho$ (dashed line) and the circle $\mathcal{C}$ in which it is distorted. Inside $\mathcal{C}$ the integral over $\rho$ is not convergent. (bottom) Deformation of the circle $\mathcal{C}$ including a bump around the origin, where the action $S''$ in Eq. (24) shows an essential singularity. The integration contour is split into $\gamma_u$ and $\gamma_d$ which are two (distorted) half-circles, above and below the real axis, respectively.

Within perturbation theory, this is sufficient to prove the same equality:

$$\langle z^n (z^*)^m\rangle = \langle x^n (iy)^m\rangle, \quad (17)$$

for any theory since the non-Gaussian part of the action, that is, the non-quadratic part of the exponential, is expanded around the Gaussian measure. Of course, the weak point of this derivation is the interchange between the series expansion and the integration.

**Contour deformations in the Doi-Peliti path-integral**

We present here a proof showing under which conditions the Doi-Peliti path-integral written in terms of complex-conjugate (coherent states) fields $\phi$ and $\phi^*$ can be converted by means of contour deformations into a path-integral written in terms of a real field $\tilde{\psi}$ and a purely imaginary field $\psi$. The proof we show here is directly inspired by appendix A of [1], where a different contour deformation is used to make $\psi$ real and $\tilde{\psi}$ purely imaginary. It is exposed for simplicity for the reversible coagulation-decoagulation process (without diffusion):

$$A \xrightarrow{\alpha_2} 2A, \quad 2A \xrightarrow{\beta_1} A. \tag{18}$$

and is generalized below to the case with diffusion. It can also be adapted to all the models considered in the main text.

The corresponding generating functional of correlation and response functions reads:

$$\mathcal{Z} = \int_{\mathbb{C}} \frac{d\phi \, d\phi^*}{2i\pi} \, e^{-S[\phi,\phi^*]+J\phi+\tilde{J}\phi^*} \tag{19}$$

where

$$S[\phi,\phi^*] = -\alpha_2 \left((\phi^*)^2 - \phi^*\right)\phi - \beta_1 \left(\phi^* - (\phi^*)^2\right)\phi^2 \tag{20}$$

and we recall that $d\phi \, d\phi^* = d\phi_1 \, d\phi_2$ with $\phi = (\phi_1 + i\phi_2)/\sqrt{2}$, and that $\alpha_2$ and $\beta_1$ are positive since they are reactions rates. Notice that for clarity we first consider that the time is frozen, and furthermore that $J = \tilde{J} = 0$, the case with sources and with time-dependence will be treated below.

**Step 1.** The proof begins with a first change of variables and the introduction of polar coordinates:

$$\phi = \rho e^{i\theta}, \quad \phi^* = \rho e^{-i\theta}, \tag{21}$$

such that we now have:

$$\mathcal{Z} = \int_0^{\infty} \frac{2\rho \, d\rho}{2i\pi} \int_0^{2\pi} d\theta \, e^{-S'[\rho,\theta]}. \tag{22}$$

**Step 2.** When $\rho$ is fixed, the integration over the angle $\theta$ can be rewritten as a contour integration on the circle of radius $\rho$ over the variable $\tilde{\psi} = \rho e^{-i\theta}$:

$$\mathcal{Z} = \int_0^{\infty} \frac{2\rho \, d\rho}{2\pi} \oint_{|\tilde{\psi}|=\rho} \frac{d\tilde{\psi}}{\tilde{\psi}} \, e^{-S''[\rho^2,\tilde{\psi}]} \tag{23}$$

with

$$S''[\rho^2,\tilde{\psi}] = \beta_1 \rho^4 \left(1 - \frac{1}{\tilde{\psi}}\right) + \alpha_2 \rho^2 (1 - \tilde{\psi}). \tag{24}$$

Since at fixed $\rho$ the integrand is holomorphic on $\mathbb{C}^*$ and in anticipation of step 4, we distort the integration contour over $\tilde{\psi}$ on the smallest contour $\mathcal{C}$ such that the integral over $\rho$ remains convergent at large $\rho$, which is given by the locus of points where the coefficient in front of $\rho^4$ in the action (24) has a positive real part:

$$\text{Re}\left(1 - 1/\tilde{\psi}\right) > 0 \quad \Leftrightarrow \quad |\tilde{\psi}| > \cos(\arg(\tilde{\psi})), \tag{25}$$

which defines a circle, see Fig. 1a.

**Step 3.** The contour over $\tilde{\psi}$ is now deformed, for all $\rho$, into the circle $\mathcal{C}$ previously defined. Given that the integrand in Eq. (23) has an essential singularity at $\tilde{\psi} = 0$, $\mathcal{C}$ is in fact slightly distorted to keep the singularity inside the integration domain, see Fig. 1b.

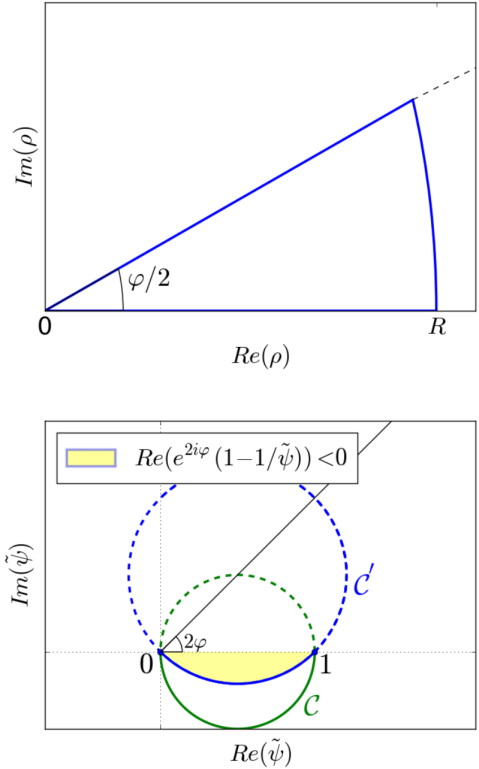

FIG. 2. (top) Wick rotation by angle $\varphi$ of the integration axis in the complex $\rho$-plane. The contribution of the integral over the circular arc of radius $R$ vanishes when $R \to \infty$. (bottom) In the complex $\tilde{\psi}$-plane, the circle $\mathcal{C}$ defined by the convergence condition (25) is modified by the Wick rotation and becomes the circle $\mathcal{C}'$. The forbidden region (shaded area) is also modified in consequence.

Remark that along this particular contour, the real part of the leading term in $\rho^4$ of the action (24) vanishes, and the convergence at large $\rho$ of the integral is now determined by the sub-leading term in $\rho^2$. The convergence is therefore guaranteed whenever

$$\text{Re}\left(\alpha_2 \left(1 - \tilde{\psi}\right)\right) > 0, \tag{26}$$

which is verified on the half-plane $\text{Re}(\tilde{\psi}) < 1$ since $\alpha_2 > 0$, and in particular all along the previously defined distorted circle $\gamma_u \cup \gamma_d$, see Fig. 1b. The partition function can therefore be rewritten as:

$$\mathcal{Z} = \mathcal{Z}_u + \mathcal{Z}_d = \int_0^{\infty} \frac{2\rho \, d\rho}{2\pi} \left(\int_{\gamma_u} + \int_{\gamma_d}\right) \frac{d\tilde{\psi}}{\tilde{\psi}} \, e^{-S''[\rho^2,\tilde{\psi}]}. \tag{27}$$

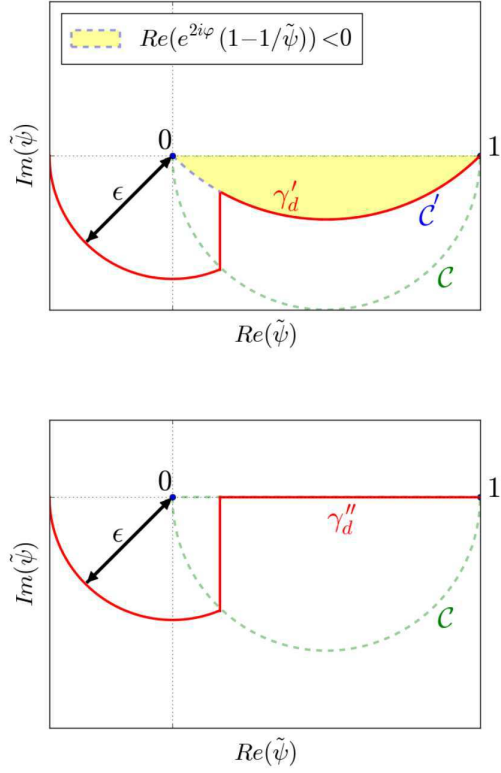

FIG. 3. (top) New contour $\gamma_d'$ obtained after the Wick rotation. (bottom) Contour $\gamma_d''$ obtained by taking the limit $\varphi \to \pi/4$ of $\gamma_d'$.

**Step 4.** We first consider the integration over the lower contour $\gamma_d$. The integration over $\rho$ is Wick-rotated by an angle $\varphi$ which means that the integration axis is tilted by an angle $\varphi$ in the complex $\rho$-plane. This is allowed because no singularity is swept by the integration axis during the rotation and because the integration at infinity does not contribute, see Fig. 2a. Then, $\mathcal{Z}_d$ reads:

$$\mathcal{Z}_d = \int_0^{\infty e^{i\varphi}} \frac{2\rho \, d\rho}{2\pi} \int_{\gamma_d} \frac{d\tilde{\psi}}{\tilde{\psi}} e^{-S''[\rho^2, \tilde{\psi}]} \qquad (28)$$

and the change of variable $\rho^2 = \rho'^2 e^{i\varphi}$ yields:

$$\mathcal{Z}_d = \int_0^{\infty} \frac{2\rho' \, d\rho'}{2\pi} e^{i\varphi} \int_{\gamma_d} \frac{d\tilde{\psi}}{\tilde{\psi}} e^{-S''[\rho'^2 e^{i\varphi}, \tilde{\psi}]}. \qquad (29)$$

**Step 5.** After the Wick rotation, the convergence condition (25) becomes:

$$\mathrm{Re}\left(e^{2i\varphi}\left(1 - \frac{1}{\tilde{\psi}}\right)\right) > 0, \qquad (30)$$

which defines a new circle $\mathcal{C}'$ (see Fig. 2b), and the contour $\gamma_d$ is modified into a new one $\gamma_d'$ which follows $\mathcal{C}'$, except for a small detour in order to avoid the singularity, see Fig. 3a.

**Step 6.** In the limit $\varphi \to \pi/4$, the radius of the circle $\mathcal{C}'$ goes to infinity and the lower part of the circle that we are considering becomes the interval $[0, 1]$ and the integration contour becomes $\gamma_d''$, see Fig. 3b:

$$\mathcal{Z}_d = \int_0^{\infty} \frac{2\rho' \, d\rho'}{2\pi} e^{i\pi/4} \int_{\gamma_d''} \frac{d\tilde{\psi}}{\tilde{\psi}} e^{-S''[\rho'^2 e^{i\pi/4}, \tilde{\psi}]}. \qquad (31)$$

The limit $\epsilon \to 0$ is still singular and is therefore not performed yet.

**Step 7.** A new Wick rotation of angle $\pi/4$ is performed on $\rho$, as well as a new change of variable $\rho'^2 = \rho''^2 e^{i\pi/4}$:

$$\mathcal{Z}_d = \int_0^{\infty} \frac{2\rho'' \, d\rho''}{2\pi} \int_{\gamma_d''} \frac{d\tilde{\psi}}{i\tilde{\psi}} e^{-S''[i\rho''^2, \tilde{\psi}]}. \qquad (32)$$

**Step 8.** A last change of variable $\psi = i\rho''^2/\tilde{\psi}$ performed to get back to cartesian coordinates finally yields:

$$\mathcal{Z}_d = \int_0^{i\infty} \frac{d\psi}{2\pi} \int_{\gamma_d''} d\tilde{\psi} \, e^{-S[\psi, \tilde{\psi}]} \qquad (33)$$

where $S$ is indeed the very same action as the one we started from in Eq. (20). The limit $\epsilon \to 0$ is now trivial because there is no more singularity when $\tilde{\psi} \to 0$ (the singularity was an artifact coming from the polar variables). Therefore, the integration over the path $\gamma_d''$ is simply an integration over the segment $[0, 1]$, and $\mathcal{Z}_d$ finally reads:

$$\mathcal{Z}_d = \int_0^{i\infty} \frac{d\psi}{2\pi} \int_0^1 d\tilde{\psi} \, e^{-S[\psi, \tilde{\psi}]} \qquad (34)$$

The computation of $\mathcal{Z}_u$ follows the same steps as those for $\mathcal{Z}_d$ up to the difference that the lower half-plane is replaced by the upper one. We finally get the expected result, that is, two complex conjugated variables are replaced by two independent variables, one purely imaginary, the other real. The important subtlety is that the real variable $\tilde{\psi}$ is compact with the proper integration boundaries given by the contour deformations, which is *a priori* far from being trivial:

$$\mathcal{Z} = \int_{-i\infty}^{i\infty} \frac{d\psi}{2\pi} \int_0^1 d\tilde{\psi} \, e^{-S[\psi, \tilde{\psi}]} \qquad (35)$$

The quadratic term in $\psi$ of the action $S$, Eq. (20), can now be eliminated by a Hubbard-Stratonovich transformation:

$$\exp\left[\beta_1\left(\tilde{\psi} - \tilde{\psi}^2\right)\psi^2\right] = \int d\eta \exp\left[-\left(\eta^2/2 + \sqrt{2\beta_1\left(\tilde{\psi} - \tilde{\psi}^2\right)}\psi\eta\right)\right]. \qquad (36)$$

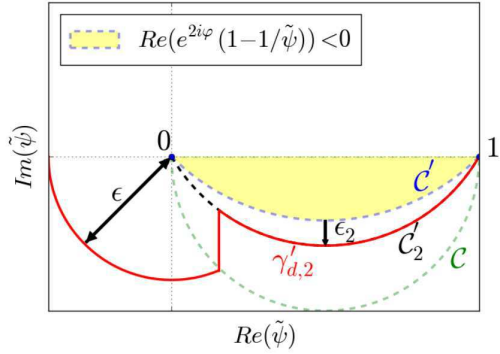

FIG. 4. Taking into consideration time and spatial dependence: a new contour $\gamma'_{d,2}$, shifted from the contour $\gamma'_d$ by a factor $\epsilon_2$, is defined such that the sub-leading terms are never relevant.

If we now reintroduce the time-dependence in $S$ by adding the term $\psi \partial_t \tilde{\psi}$, the integration over the imaginary variable $\psi$ in Eq. (35) yields a Dirac-delta function of the LE over the real variable $\tilde{\psi}$:

$$\partial_t \tilde{\psi} = \alpha_2 \left( \tilde{\psi}^2 - \tilde{\psi} \right) + \sqrt{2\beta_1 \left( \tilde{\psi} - \tilde{\psi}^2 \right)} \, \eta \qquad (37)$$

where $\eta$ is in fact a Gaussian white noise since its probability distribution is given in Eq. (36) by $\mathcal{P}(\eta) = \exp\left( -\eta^2/2 \right)$.

Remark that taking into account the time-dependence does not invalidate the proof because the extra $\psi \partial_t \tilde{\psi}$ term only adds a sub-leading term (in factor of $\rho^2$) to the action $S''$ defined in Eq. (24). This sub-leading term is significant only when the contour is exactly on the circle $\mathcal{C}'$. One can therefore slightly shift all the previously defined contours by a factor $\epsilon_2$ in order not to lie on the circle $\mathcal{C}'$, as illustrated on Fig. 4 in the case of the contour $\gamma'_d$. The rest of the proof remains unchanged, except that the limit $\epsilon_2 \to 0$, which is not singular, has to be taken at the end.

The spatially-extended case is treated in the same way since the extra term coming from spatialization also contributes only in a sub-leading way.

Let us now briefly discuss the case with sources, with space and time frozen. In the limit where the sources $J$ and $\tilde{J}$ are infinitesimal, the original partition function (19) can be expanded as a power series and reads:

$$\begin{aligned}
\mathcal{Z}[J, \tilde{J}] &= \int \frac{\mathrm{d}\phi \, \mathrm{d}\phi^*}{2i\pi} \, e^{-S[\phi, \phi^*] + J\phi + \tilde{J}\phi^*} \\
&= \int \frac{\mathrm{d}\phi \, \mathrm{d}\phi^*}{2i\pi} \, e^{-S[\phi, \phi^*]} \left( 1 + J\phi + \frac{1}{2}J^2\phi^2 + \cdots \right) \times \\
&\quad \left( 1 + \tilde{J}\phi^* + \frac{1}{2}\tilde{J}^2(\phi^*)^2 + \cdots \right)
\end{aligned} \qquad (38)$$

and the deformations described above can be applied to all the terms in the sum, which proves that adding infinitesimal sources is not an issue.

Notice that the proof was shown here for the particular choice of reactions (18) but is in fact generic for any set of reactions of the form $A \to pA$, $2A \to qA$ with $q < 2$, except that the deformation contours would be different. For instance, if one were to consider the pure annihilation reaction $2A \to \emptyset$, the convergence condition Eq. (25) would be modified and would define a lemniscate instead of the circle $\mathcal{C}$.

**Spatially-extended case.** In the field-theoretic formulation, we showed in the previous section that the spatially-extended case is handled by a slight deformation of the contours in order to take care of the sub-leading term coming from diffusion. We describe here how it is treated in the probability-generating function formalism.

To take into account the spatial structure in the RD process, the system is now described by a $d$-dimensional lattice indexed by an integer $i$. Each site has a number $N_i(t)$ of particles at time $t$. A hopping reaction between nearest neighbours is introduced to account for diffusion:

$$A\emptyset \overset{D}{\leftrightarrow} \emptyset A \qquad (39)$$

where $D$ is the diffusion coefficient. The state of the system is now described by $P_{\{n\}}(t) \equiv \mathrm{Prob}(N_1(t) = n_1, \ldots, N_m(t) = n_m)$, whose time-evolution is still described by the master equation with the additional diffusion term:

$$\frac{D}{h^2} \sum_{<i,j>} \left( (n_i + 1) P_{\ldots, n_i-1, n_j+1, \ldots} - n_i P_{\{n\}} \right) \qquad (40)$$

where $<i,j>$ means that sites $i$ and $j$ are nearest neighbours, and $h$ is the lattice spacing. In this context, the probability-generating function for a spatially-extended system now reads:

$$G(z_1, \ldots, z_n, t) = \left\langle z_1^{N_1(t)} \cdots z_n^{N_m(t)} \right\rangle_{\mathrm{RD}} \qquad (41)$$

and is subject to the Fokker-Planck equation:

$$\begin{aligned}
\partial_t G &= \frac{D}{h^2} \sum_{<i,j>} (z_j - z_i) \partial_{z_i} G \\
&\quad + \sum_i A(z_i) \partial_{z_i} G + \sum_i B(z_i) \partial_{z_i}^2 G
\end{aligned} \qquad (42)$$

where the functions $A$ and $B$ are defined in the main text. One notices that the term coming from the diffusion defines a discrete Laplacian over $z_{\{n\}}$. After taking the continuous limit in space ($z_{\{n\}} \to z(x)$), it therefore yields the following LE for the spatialized stochastic variable $Z \equiv Z(x,t)$:

$$\partial_t Z = D\nabla^2 Z + A(Z) + \sqrt{2B(Z)} \, \eta \qquad (43)$$

where $\eta \equiv \eta(x,t)$ is a Gaussian white noise.